\documentclass[sigchi]{acmart}
\usepackage[T1]{fontenc}
\usepackage[utf8]{inputenc}

\graphicspath{{img/}}

\usepackage{lipsum}
\usepackage{graphicx}
\usepackage{amsmath}
\usepackage{fullpage}
\usepackage{mathptmx} 
\usepackage{helvet} 
\usepackage[utf8]{inputenc}
\usepackage{relsize}
\usepackage[T1]{fontenc}
\usepackage{tikz}
\usepackage{booktabs}
\usepackage{textcomp}
\usepackage{multirow}
\usepackage{pgfplots}
\usepackage{url}
\usepackage{footnote}
\usepackage{subcaption}

\usepackage[french,english]{babel}
\newcommand{\fr}[1]{\emph{\foreignlanguage{english}{#1}}}

\copyrightyear{}
\acmYear{}
\setcopyright{acmCopyright}
\acmConference[Master Thesis for Master of Science in Informatics at Grenoble, Ubiquitous and Interactive Systems]{Master of Science in Informatics at Grenoble}{December, 2019}{Grenoble, France}
\acmBooktitle{IHM'19~: 31\textsuperscript{ème} conférence Francophone sur l'Interaction Homme-Machine, December 10--13, 2019, Grenoble, France}
\acmPrice{}
\acmDOI{}
\acmISBN{}

\begin{document}
\selectlanguage{french}
\sloppy

%
\title[Interactions par franchissement grâce a un système de suivi du regard]{Interactions par franchissement grâce a un système de suivi du regard}
\subtitle{\fr{Investigation of Crossing Interaction with Gaze-Tracking}}

\author{Sébastien RIOU, Didier SCHWAB, François BERARD}
\affiliation{%
 \institution{Univ. Grenoble Alpes, Grenoble-INP, LIG}
 \city{38000 Grenoble}
 \state{France}
}
\email{{sebastien.riou, didier.schwab, francois.berard}@univ-grenoble-alpes.fr}

%
%

%

%
\begin{abstract}

Human-computer interactions  based on gaze-tracking have spread during the last few years. Video games, applications in health, trading, market research, and many other fields have started to use this new technology that seems {\em invisible} to the user. However, the dominant form of interaction using gaze tracking uses dwell-time for command activation; which introduces strong constraints in the interaction: dwell-time activation requires users to look steadily at an element for a predefined amount of time in to select it. While dwell-time alleviates a part of the Midas touch problem (referring to the fact that an element fixed by the user will be activated even if it was not intended to do so), it doesn't completely remove it: users should not gaze too long on an item or they may trigger an unintended activation. In addition, dwell-time slows down users  interaction by requiring a pause each time an activation is needed. In this project, we study an alternative selection method based on crossing interactions, a well-studied method used in conventional HCI. This interaction allows users' gaze to rest in areas that don't have crossing triggers, and it removes the need to pause in the interaction. We found that crossing interaction had similar performances than dwell-time interaction with novice users. The performance was even better for users having previous experience with gaze interaction.
 \end{abstract}

%
%


%

%

%
\maketitle


\section*{\fr{Résumé}}
\fr{
Les interactions homme-machine basées sur le suivi du regard se sont étendues au cours des dernières années. Les jeux vidéo, la santé, le commerce, les études de marché et bien d'autres domaines ont commencé à utiliser cette nouvelle technologie qui semble {\em invisible} à l'utilisateur. Cependant, la forme d'interaction la plus courante utilisant le suivi du regard utilise le temps de fixation pour l'activation des commandes ; ce qui introduit de fortes contraintes dans l'interaction : l'activation par temps de fixation exige que l'utilisateur regarde un élément pendant un temps prédéfini pour le sélectionner. Bien que cette interaction atténue une partie du problème de toucher Midas (se référant au fait qu'un élément fixé par l'utilisateur sera activé même s'il n'en avait pas l'intention) il ne le supprime pas complètement : les utilisateurs ne doivent pas regarder trop longtemps sur un élément ou ils peuvent déclencher une activation involontaire. De plus, le temps de fixation ralentit l'interaction de l'utilisateur en exigeant une pause chaque fois qu'une activation est nécessaire. Dans ce projet, nous étudions une méthode de sélection alternative basée sur les interactions par franchissement, une méthode déjà étudiée et utilisée en IHM conventionnelle. Cette interaction permet aux utilisateurs de  reposer le regard dans des endroits sans zone de franchissement et élimine le besoin de faire une pause dans l'interaction. Nous avons constaté que l'interaction par franchissement avait des performances similaires à celles temps de fixation avec des utilisateurs novices. La performance était encore meilleure pour les utilisateurs ayant une expérience antérieure de l'interaction avec le regard.
%
}

\section*{\fr{Mots Clés}}
\fr{Interaction par franchissement, oculométrie}


\section{Introduction}

Les interactions par oculométrie ont été développées et se sont multipliées au cours des dernières années. Les jeux vidéo, la santé, le commerce, les études de marché et de nombreux autres domaines ont commencé à se tourner vers cette nouvelle technologie transparente pour les utilisateurs. Malheureusement les interactions actuellement proposées par le biais de ces appareils sont vraiment limitées et contraignantes. En effet, seule la méthode par temps de fixation (\textit{dwell-time interaction}), visant à sélectionner un élément à l'écran après l'avoir regardé durant un certain temps, est généralement utilisée.

 Cette technique a l'avantage d'éviter le problème connu sous le nom de \textit{toucher de Midas}. Ce nom fait référence à Midas, le roi de Phrygie, un pays de l'Asie Mineure antique, qui, selon la légende, obtint le vœux de transformer en or tout ce qu'il touchait.  Par analogie, dans le cas de oculométrie,  le problème consiste au fait qu'un élément fixé par l'utilisateur pourrait être activé en intéraction directe même si ce n'était pas dans son intention. , elle ralentit toujours grandement l'action et l'apprentissage de l'utilisateur, s'avérant rapidement très frustrante. 

L'objectif de cette étude est de proposer et de tester une méthode alternative d'interaction sans temporisation permettant d'éviter le problème du toucher de Midas  qui soit plus rapide que la méthode par fixation habituellement utilisée. Il s'agit de l'interaction par franchissement (\textit{crossing interaction}), introduite par Accot en 2002 \cite{accot2002}. Son principe général consiste à activer un élément en traversant une frontière. En particulier, il n'exige pas l'utilisation d'une autre modalité (comme l'appui sur un bouton) pour la validation. 


\textbf{Contributions.} Les principales contributions de cet article sont les suivantes :
\begin{itemize}
\item un l'état de l'art de différentes études sur la méthode de franchissement qui pourrait être intéressant dans l'adaptation pour les systèmes de suivi oculaire ;
\item nous avons proposé une formule pour calculer le nombre maximal de tranches qui peuvent être placées dans un menu circulaire utilisable pour les eye-trackers.
\item nous avons adapté la méthode d'interaction de franchissement au suivi oculaire ;
\item nous avons testé l'efficacité de l'interaction de franchissement avec un eye-tracker à bas coût (Tobii 4C, $\approx$170€) avec des personnes qui ont déjà utilisé un dispositif de suivi oculaire et des personnes qui n'en ont jamais utilisé.
\end{itemize}

%
%

\textbf{PRÉSENTATION DE L'ARTICLE}

\section{Travaux liés}

L'interaction par franchissement consiste à traverser un élément afin de déclencher une action. Ainsi, traverser une zone permet d'activer le bouton correspondant. 


\begin{figure}[!h]
    \center
    \includegraphics[width=0.7\columnwidth]{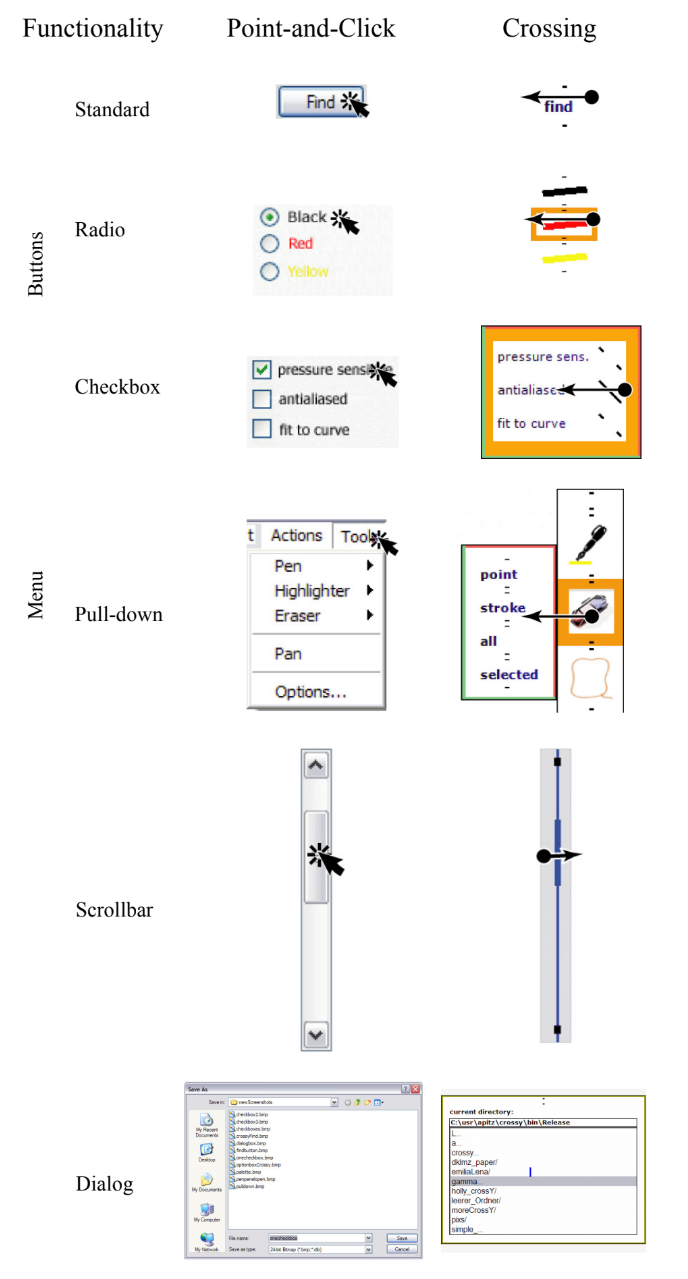}
    \caption{{\em The correspondence table showing elements of traditional, point-and-click GUIs and their CrossY counterparts.} by Apitz \cite{apitz} }
    \label{crossY}
\end{figure}

L'étude d'Accot \cite{accot2002} compare l'interaction traditionnelle par pointer-cliquer avec l'interaction par franchissement en montrant que chacun d'eux suivait une loi de Fitts. Cette étude a montré que le franchissement et le pointer-cliquer avaient des paramètres de Fitts d'une valeur similaire pour le même type de tâche. Le franchissement a finalement été considéré comme aussi efficace qu'un simple pointer-cliquer sur l'interface WIMP classique. Il peut même être intéressant en toucher direct, comme le montrent Luo et al. \cite{luo2014}. Les interactions directes sont généralement naturelles pour l'utilisateur et peuvent être utiles pour réduire le temps nécessaire à l'activation d'un objet, car il n'y a pas de dispositif intermédiaire. Cao et al. \cite{cao2006} expliquent également dans leur étude que les techniques de sélection par franchissement peuvent fournir une interaction rapide et fluide avec une interface basée sur un projecteur manuel.


Ce principe de franchissement est bien décrit dans l'étude crossY d'Apitz et coll. \cite{apitz}. La figure \ref{crossY} compare la manière traditionnelle et la manière de sélections de crossY. Apitz voulait comparer l'efficacité de l'interaction de franchissement avec le pointer-et-cliquer par stylets. L'idée était d'éviter l'alternance entre un mode de saisie très naturel et fluide pour esquisser ou prendre des notes et une interaction très rigide et segmentée en utilisant les éléments de l'interface graphique. Le franchissement conduit finalement à une composition fluide de commandes ce qui est parfaitement adapté à notre interaction oculaire dans laquelle nous voulons utiliser la saccade à la place de la fixation car cette dernière n'est pas quelque chose de naturel pour un être humain.


L'interaction par franchissement a déjà été utilisée pour les personnes en situation de handicap moteur et son efficacité a été testée avec différentes modalités. Wobbrock \cite{wobbrock2007,wobbrock2008} montre dans ses articles que pour les personnes qui interagissent avec le trackball, par exemple, il est plus facile et plus rapide de gérer une interaction par franchissement par rapport à une interaction pointer-cliquer qui nécessite plus de sous-actions et que l'utilisateur doit vraiment être précis. 
Le franchissement par oculomètre a également été essayé, principalement pour les logiciels de saisie textuelle. C'est le cas, par exemple, dans le système de saisie de Kurauchi \cite{kurauchi2016,kurauchi2018}, car il évite les temps de fixation.
Cependant, cette dernière étude ne peut être appliquée que dans ce champ très restreint de la saisie de texte, par rapport à notre objectif qui serait adapté à différentes situations (différents menus de sélection par exemple).


\section{Contributions}
\subsection{Description du menu par franchissement}

Notre première idée, pour la création de ce menu par franchissement, a été d'étudier la conception de menus circulaire (\emph{marking menus)} exploitables par des eye-trackers. Dans un premier temps, nous avons visualisé le tracé de chemins réalisés par le regard. Cette expérience nous a permis de constater qu'un menu circulaire est inutilisable avec le regard : en effet, il n'est pas dans la nature de l'œil humain de suivre un chemin homogène avec les yeux. 


Nous avons choisi un menu circulaire qui parait être une approche viable pour une interaction par franchissement du regard. La forme circulaire du menu est plus adaptée que le menu grille, puisque la répartition des éléments sur le menu grille aurait conduit à déclencher une action à chaque fois que l'utilisateur passe d'un élément à un autre (en traversant une ligne ou une rangée du menu). En d'autres termes, un menu de grille créerait un problème similaire au toucher de Midas rendant le menu impossible à utiliser. Pour la mise en œuvre de notre interaction, nous nous sommes inspirés du menu pEye conçu par Huckauf \cite{huckauf2007,huckauf20081} (cf fig \ref{pEyeA}). Cette interaction a été adaptée aux mouvements du regard humain et à nos attentes en matière d'interaction pour la sélection. 


La première chose à faire est de calculer le nombre maximal d'éléments qui peuvent être placés dans ce menu. Dans le travail de \cite{urbina2010}, un menu à 6 tranches est considéré comme "optimal" sans tenir compte de la largeur du menu ce qui a pourtant semblé être l'une des variables les plus importantes à explorer. Nous avons ainsi commencé à calculer un nombre de tranches plus précis, en fonction de la dimension du menu et d'autres contraintes importantes comme le modèle eye-tracker. 


La conception finale utilisée lors de la phase de tests utilisateurs est donc un menu circulaire où chaque section contient un élément. Pour sélectionner un élément, l'objectif est d'obtenir une interaction de franchissement en déplaçant notre regard de la section vers l'extérieur du menu. L'événement de franchissement sera déclenché entre les limites de l'élément de menu et une distance proche du menu.


\subsection{Hypothèses}

Dans cette section, nous examinons les hypothèses que nous avons prises pour le calcul du nombre d'éléments du menu.

\subsubsection{Permettre une exploration facile}

L'interaction de franchissement dans notre menu doit se déclencher dès que le regard sort du menu. Cela signifie qu'à l'intérieur du menu, l'utilisateur doit pouvoir explorer librement les éléments, sans avoir un retour visuel comme avec la barre de progression circulaire de l'interaction par fixation et sans craindre une activation non souhaitée.  Par rapport à la méthode du dwell-time, la méthode du croisement crée une "pause" (break) avant la sélection, et non pendant la sélection (alors que l'utilisateur ne doit pas bouger pendant le dwell-time). L'utilisateur cherche d'abord l'élément qu'il souhaite sélectionner, puis il le sélectionne.  Ces deux actions peuvent se faire séparément ou d'une traite contrairement à l'interaction par fixation qui doit toujours se faire en deux étapes. 


\subsubsection{Plus rapide et moins fatigant que l'interaction par fixation}

L'œil humain est constamment en mouvement par des saccades ou des micro-saccades. Ainsi, l'utilisation de l'interaction par fixation peut être difficile à gérer sur de petits objets puisque les yeux peuvent, par exemple, bouger "involontairement" à cause des micro-saccades ou ils peuvent être attirés par quelque chose dans la vue périphérique.


Le fait de transformer l'action en une interaction par franchissement est plus convivial pour l'utilisateur car l'humain utilise de manière naturelle plusieurs fois par minute des saccades.
La saccade oculaire est ainsi à la fois plus précise et moins fatigante que la fixation. L'action demandée à l'utilisateur peut devenir moins fatigante parce que la sélection d'une cible par interaction par franchissement ne demande pas nécessairement à l'utilisateur d'être aussi précis que pour le pointage comme Wobbrock l'a montré avec sa comparaison franchissement/pointage avec souris et trackball \cite{wobbrock2007}.
De plus, lorsque l'interaction par franchissement est utilisée en mode expert, nous pensons qu'une grande partie des 500 millisecondes habituellement dépensées par le temps de fixation peut être économisée lors de l'activation, ce qui permet une action plus rapide.


\subsubsection{Taille et nombre de sections du menu circulaire} \label{computation}

L'oculomètre utilisé dans cette étude, le 4C de la société suédoise Tobii est considéré par le fabriquant comme bien calibré si 93 \% des résultats de la population d'utilisateurs a un décalage maximum de 0,8 degré en justess et 0,5 degré en précision{Tobii{Tobii {\em Justesse et méthode de test de précision pour eye trackers distants} : https://www.tobiipro.com/learn-and-support/learn/eye-tracking-essentials/}, c'est-à-dire qu'il a un total de 1,3 degrés d'incertitude lorsqu'il est utilisé à une distance de 50 à 95 cm pour l'utilisateur donnée dans les spécifications du Tobii 4C \footnote {Tobii 4C description page : https://gaming.tobii.com/product/tobii-eye-tracker-4c/}. 


Considérons d'abord une distance de lecture confortable, généralement comprise entre 38,1 et 63,5 cm, avec un texte écrit en 12 points ($\approx$ 0,45 cm).


Nous voulons afficher une lettre à l'intérieur d'un cercle ou d'un carré. Pour le rendre plus lisible, nous ajoutons une petite marge entre les lettres et le bord de son conteneur, pour une lettre de taille 12pts, la marge sera également égale à 12pts comme indiqué sur la figure \ref{LetterSize}.


\begin{figure}[!h]
    \center
    \includegraphics[width=0.7\columnwidth]{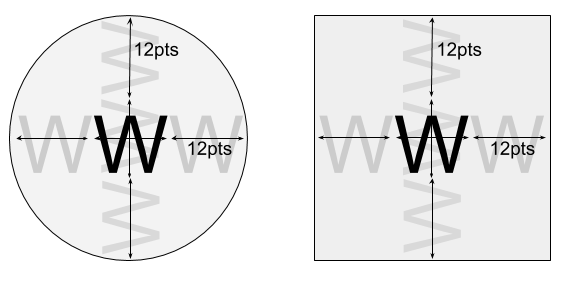}
    \caption{Position et taille de la lettre dans la boîte circulaire et rectangulaire}
    \label{LetterSize}
\end{figure}

En considérant la distance confortable moyenne entre 38,1 et 63,5 cm et en connaissant la distance de travail de l'oculomètre, on considère une distance confortable entre 50 et 63,5 cm. En utilisant la fig \ref{angle} nous pouvons voir que l'angle est environ :


$$ angle = 2 \times \frac{angle}{2} $$
$$= 2 \times atan(\frac{\frac{3 \times TailleCharact\grave{e}re}{2}}{DistanceLectureConfortable})$$
$$ = 2 \times atan(\frac{\frac{3 \times 0.45}{2}}{\frac{50+63.5}{2}}) \approx 1,36 \mbox{\degre} $$

\begin{figure}[!h]  
    \center 
    \includegraphics[width=0.7\columnwidth]{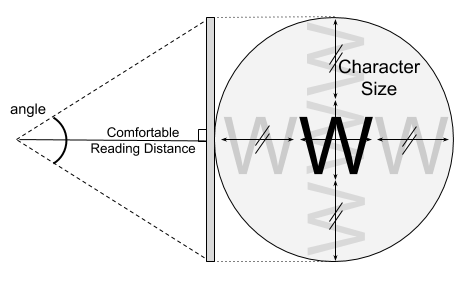}
    \caption{Calcul de l'angle de vision en fonction de la distance de lecture et d'une taille confortables}
    \label{angle}
\end{figure}

Évidemment, l'angle de vision confortable est plus grand que l'angle d'incertitude fourni par Tobii (1,36\degre >1,3\degre). Par conséquent, nous pouvons voir que la limitation de notre conception en terme de taille de l'élément doit être donnée par le confort de l'œil humain lui-même.


Maintenant que nous avons l'angle de distance pour un objet, considérons un utilisateur avec 10/10 corrigé sur chaque œil à une distance de 50 à 60 cm de l'oculomètre. Il est possible de calculer le nombre maximal d'éléments pouvant composer un menu grâce aux dimensions de ce menu, au degré maximal d'incertitude donné par la limitation du confort oculaire, on peut déduire la taille minimale d'un objet sélectionnable à l'écran.


Considérons d'abord un menu circulaire avec un rayon R, le périmètre du rayon est


\[perim = 2 \times \pi \times R\]

La taille minimale de l'objet à une distance D de l'utilisateur peut être calculée géométriquement comme schématisé dans la figure \ref{objSize} :


\[objSize = 2\times D\times tan(\frac{angle}{2}) \]

\begin{figure}[!h]  
    \center 
    \includegraphics[width=0.7\columnwidth]{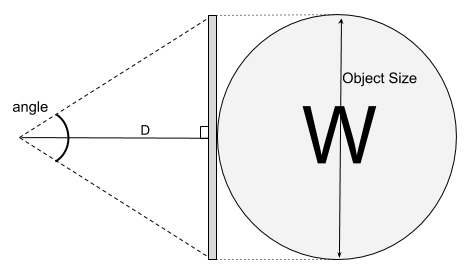}
    \caption{Calcul de la taille minimale d'un objet à la distance D suivant un certain angle de vision}
    \label{objSize}
\end{figure}

La figure \ref{sectionNumberC} montre que pour un utilisateur à une distance D d'un menu circulaire de rayon R, on obtient un nombre maximal de tranches, égal à : 

\[ \theta = 2 \times asin(\frac{\frac{objSize}{2}}{R\times2})\]

\[ arcLength = \frac{perim}{360} \times 2 \times \theta \]

\[ numberOfSlices = \left \lfloor \frac {perim}{arcLength} \right \rfloor \]

\begin{figure}[!h]  
    \center 
    \includegraphics[width=\columnwidth]{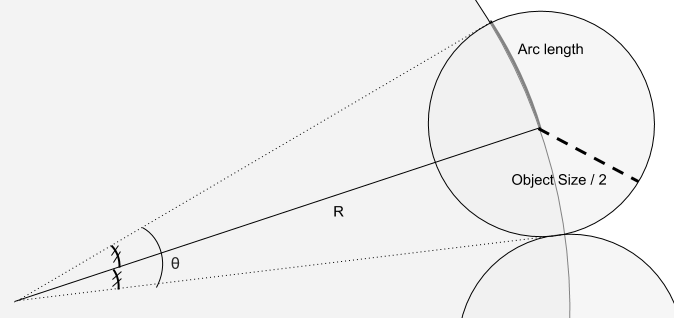}
    \caption{Calcul du nombre maximum de tranches pour les objets de taille donnée dans un menu circulaire }
    \label{sectionNumberC}
\end{figure}

 Ce calcul peut compléter l'étude de Huckauf et Urbina\cite{urbina2010} sur les limites du nombre de sections dans les menus pEYEs en proposant une formule prenant en compte le rayon du menu, la distance à l'utilisateur et l'angle utilisé comme justesse spatiale, et donnant le nombre maximal de sections qui peuvent être représentées dans un menu.
 

Maintenant, afin de concevoir une étude comparative avec les menus basés sur l'interface \textit{status quo} (interface par fixation), nous devons également calculer la taille d'un menu de grille pour un nombre donné d'éléments, afin de pouvoir l'implémenter. Nous avons décidé d'utiliser des éléments carrés et de prendre un dessin où le nombre de lignes est égal à un tiers du nombre de colonnes ce qui conduit au fait que la largeur de la grille est égale à 3 fois la hauteur (cf fig \ref{sectionNumberL}). Les dimensions de l'objet pour un objet carré sont alors égales à $objSize \times objSize$. Cela signifie que pour un menu de Largeur $W$ nous avons un nombre maximum d'objets égal à :


\[ NombredElements = \left \lfloor\frac{W}{objSize}\right \rfloor \times  \left \lfloor\frac{\left \lfloor\frac{W}{objSize}\right \rfloor}{3}\right \rfloor \]

\begin{figure}[!h]  
    \center 
    \includegraphics[width=\columnwidth]{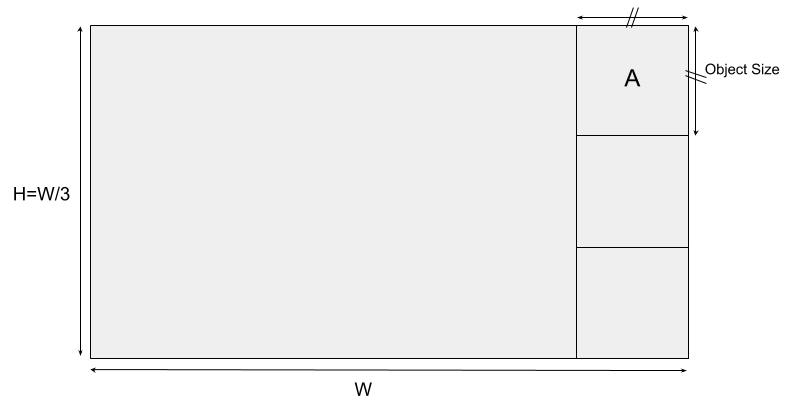}
    \caption{Calcul du nombre maximum d'éléments dans le menu grille selon une largeur donnée}
    \label{sectionNumberL}
\end{figure}

\section{Mise en place de l'expérimentation}

Nous avons conduit une expérimentation sur des utilisateurs pour observer et mesurer les bénéfices et les défauts d'une activation par le regard basée sur le principe du franchissement. 

\subsection{Déroulement de l'expérimentation}
\subsubsection{Participants}

Comme nous l'avons expliqué dans l'introduction, nous avons mené une étude initiale auprès de participants n'ayant pas de handicap. Nous avons invité 18 étudiants âgés de 21 à 33 ans et provenant de 7 nationalités différentes à participer à l'étude..

13 de ces étudiants ont participé à l'expérience "Chercher et Activer". 2 d'entre eux étaient des étudiants de Master en histoire, les 11 autres étaient des étudiants de Master en informatique, 3 d'entre eux avaient déjà utilisé un eye-tracker auparavant.

11 des 18 volontaires ont participé à l'expérience "Activation". 2 d'entre eux étaient des doctorants, les 9 autres étaient des étudiants en Master, tous issus du domaine de l'informatique. 1 des doctorants et 3 des étudiants à la maîtrise avaient déjà utilisé un traceur de regard auparavant.

7 personnes ont participé à la première expérience mais pas à la seconde. 5 autres personnes ont participé à la deuxième expérience mais pas à la première.

Pour les deux expériences, les utilisateurs ayant déjà utilisé une technologie d'oculométrie peuvent être considérés comme des utilisateurs " experts " de cette technologie.

Les données d'un participant à la première expérience ont été abandonnées parce qu'il était daltonien et donc incapable de distinguer les couleurs pour les validations " correctes " et " incorrectes ". Cela a attiré notre attention sur la nécessité de travailler à une meilleure utilisation des couleurs dans notre système.

\subsubsection{Environnement de travail}

Les expériences sont effectuées dans une pièce lumineuse sans lumière artificielle. Les participants sont assis entre 50 et 60 cm de l'oculomètre {\em Tobii 4c} placé à 75 cm du sol. Il est connecté à un processeur un  {\em MacBookPro{\copyright}  11.3 Intel Core{\copyright}  i7-4980 HQ} fonctionnant sous {\em Windows{\copyright} 7 pro 64 bits } avec 16 Go de RAM. On demande aux participants de s'asseoir confortablement et d'essayer de ne pas bouger pendant les expériences. Pour chaque participant, un premier étalonnage du suivi du regard est effectué au début de l'expérience via le logiciel {\em Tobii{\copyright} eye tracking software} de l'oculomètre {\em  Tobii{\copyright} 4C}. Il est suivi d'une autre calibration que nous avons développée. Ce second étalonnage réduit le décalage avec l'oculomètre avant chacun des différents blocs de test. Il fonctionne en compensant le décalage moyen de 5 points différents de l'écran (en haut à droite, en haut à gauche, en bas à droite, en bas à gauche et au centre).

\subsection{Design des menu pour l'expérience}

\subsubsection{Apparence générale} 

Nous avons créé deux types de menu représentant nos deux conditions expérimentales. Le premier menu dans la condition STATUS QUO est similaire à ceux utilisés dans les logiciels.

Nous remplissons chaque menu avec les 26 lettres classées par ordre alphabétique. Les participants ont ainsi une bonne idée de l'endroit où trouver chaque cible. Cela a deux objectifs :  Premièrement, le fait de proposer aux participants une cible prévisible réduit le temps de recherche nous permettant de nous concentrer sur le temps d'activation de la cible. Deuxièmement, cela a une validité écologique car dans les applications réelles, les utilisateurs développent rapidement la connaissance de la localisation des cibles. Nous avons donc préféré utiliser des lettres comme cible plutôt que des icônes qui, bien qu'elles soient plus représentatives des applications du monde réel, ne pourraient pas fournir une disposition prévisible à nos participants.

Le modèle final du menu STATUS QUO, est représentée dans la Figure \ref{dwellMenu}. Nous avons ajouté une retour visuel dynamique au menu : lorsqu'un élément est regardé par l'utilisateur, la cellule le contenant devient jaune et un cercle de progression est affiché pour fournir un retour sur le temps de fixation.
\newline

\begin{figure}[!h]  
    \center 
    \includegraphics[width=\columnwidth]{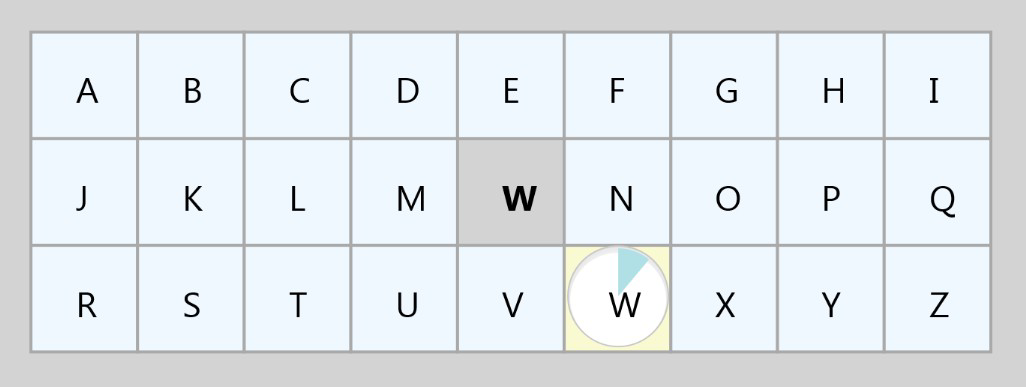}
    \caption{Dwell grid menu with the 26 letters of the latin alphabet in alphabetical order.
    The central slot will be used as text field during tests.}
    \label{dwellMenu}
\end{figure}

Le deuxième menu en condition CROSSING est le menu circulaire que nous avons créé, utilisant l'interaction par franchissement.
Nous avons remarqué pendant les séances de pré-test qu'il était difficile aux utilisateurs de dépasser intentionnellement les limites du menu pour activer la zone de franchissement (ils étaient "bloqués" à l'intérieur du menu). Nous avons donc ajouté des disques en dehors du menu, un pour chaque section, afin de créer une cible visible pour l'aider l'utilisateur à sortir du menu et traverser plus facilement la zone de franchissement. Le modèle de notre menu circulaire peut être observés sur la Figure \ref{crossingMenu} ci-dessous. Comme nous pouvons le voir, l'ajout de ces nouvelles cibles prend également de l'espace à l'écran et devra donc être pris en compte pour les implémentations futures.

\begin{figure}[h!]
\centering
  \includegraphics[width=\columnwidth]{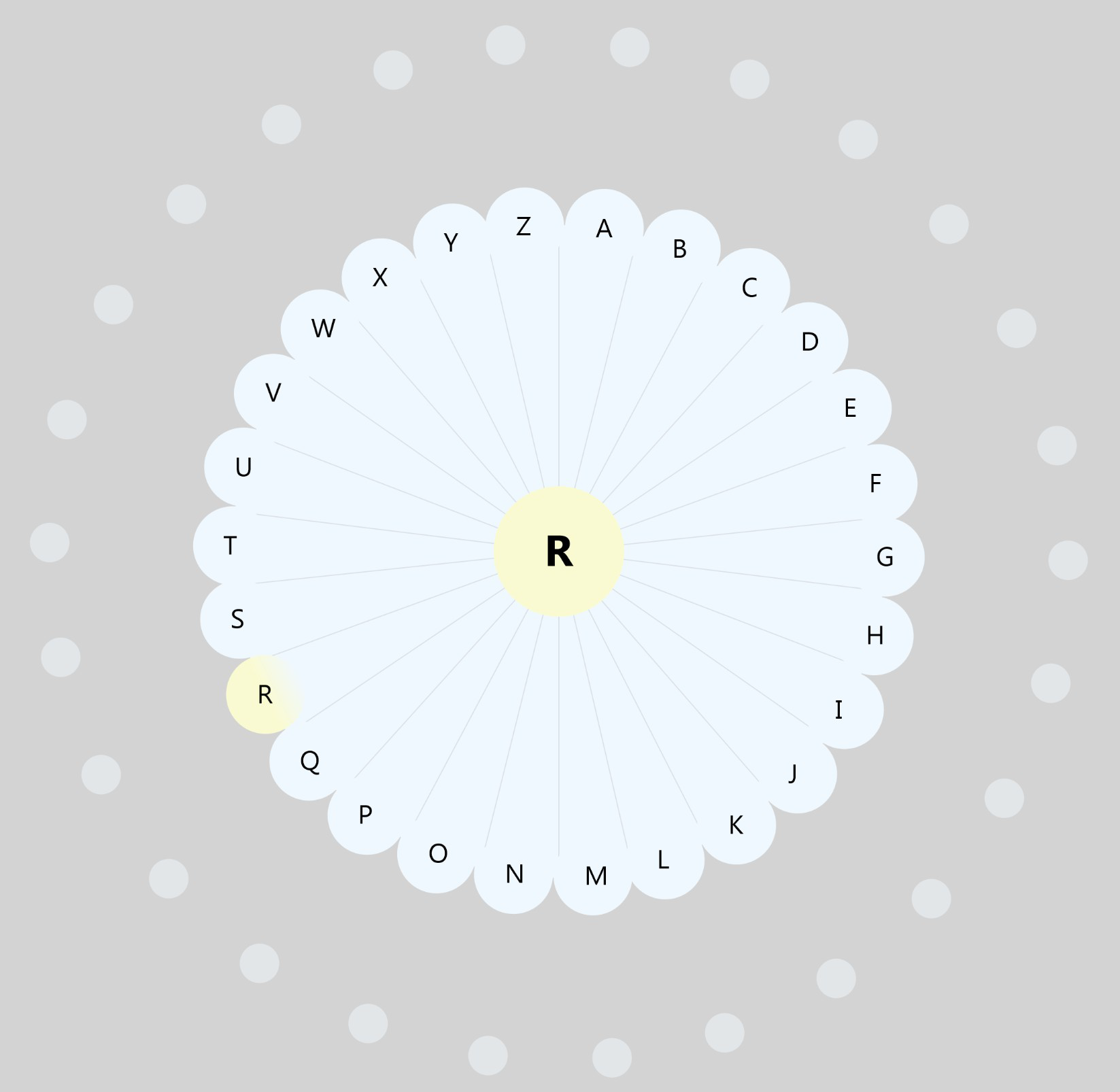}
\caption{ Main design of the circular crossing menu}
\label{crossingMenu}
\end{figure}

\subsubsection{Taille des menus}

En utilisant la taille d'un objet, calculée dans la partie \ref{computation}, pour un utilisateur assis à une distance de 50 à 63,1 cm du eye-tracker, on calcule la taille des deux menus contenant 26 éléments.

\begin{itemize}

\item Pour le menu circulaire, le rayon est de :  
\[ CrossingMenuRadius = \frac {objSize}{4 \times tan(\frac{360}{4 \times 26}) } \approx \frac {1.31}{4 \times tan(\frac{360}{4 \times 26}) } \approx 5.37 \textrm{cm}\]

\item Pour le menu en grille, les dimensions sont données par :
\[ DwellMenuWidth = 9 \times objSize \approx 9 \times 1.30 \approx 11.7cm\]
\[ DwellMenuHeight = 3 \times objSize \approx 3 \times 1.30 \approx 3.9cm \]
\[\textrm{(because} \quad 26 = 9 \times 3 -1 = (3\times 3) \times 3 - 1 \textrm{)}\]

\end{itemize}

\subsection{Taches}

\subsubsection{La Tache de {\em Recherche and Sélection}}

Nous voulons observer, pour la première tache, le temps moyen consacré par l'utilisateur à la recherche et à la sélection d'un élément donné. Dans les deux conditions STATUS QUO et CROSSING, nous affichons l'élément demandé au milieu du menu. Le participant doit le chercher dans la grille ou dans le cercle et l'activer. Nous avons demandé aux participants d'exécuter la tâche le plus rapidement possible tout en évitant les erreurs (c.-à-d. choisir une lettre différente de la lettre demandée) . Nous avons enregistré l'\verb+activation_time+ : le temps entre l'apparition de la lettre demandée et l'activation de la lettre correspondante. Pour une analyse plus poussée, nous avons également enregistré le \verb+return_time+ : le temps passé par l'utilisateur, après une activation, à revenir au centre du menu pour voir la lettre suivante.

Chaque participant effectue deux blocs du test {\em search and activate}. Les blocs sont utilisés pour offrir un temps de repos pour les yeux (entre les blocs) et pour observer un effet d'apprentissage possible pendant la séance. Pour chaque bloc, le participant doit regarder 55 lettres dans un ordre pseudo-aléatoire. Chacune des 26 lettres apparaît exactement deux fois sauf les 3 premières lettres qui apparaissent 3 fois. Ces 3 premières lettres sont utilisées comme échauffement et ne sont pas comptées dans les résultats finaux.

\subsubsection{La Tache d' {\em Activation}}


La tâche ``{\em search and activate}}'' décrite précédemment demande aux participants de rechercher différentes lettres, ce qui peut conduire à une grande variabilité dans l' \verb+activation_time+ final. Nous nous sommes donc également intéressés au ``temps d'activation brute'' des cibles, c'est-à-dire aux activations ne nécessitant pas une recherche. L'activation sans recherche a une forte validité écologique : lorsqu'on devient expert d'un système, on a souvent besoin d'activer des cibles dont on connaît l'emplacement précis.

Nous avons conçu une expérience de {\em activation} pour supprimer la majorité du temps d'{\em observation} et de {\em recherche} nécessaire à l'activation de la cible. Le participant doit activer deux lettres en alternance, avec 10 répétitions (soit 20 activations). Avant de commencer l'expérience, nous demandons aux participants de localiser les deux lettres qu'ils devront activer.
Nous faisons varier les deux lettres afin de contrôler leur distance relative, qui a un impact sur le temps d'activation. Comme dans la tâche précédente, nous utilisons les deux premières activations de chaque répétition comme échauffement et nous ne les comptons pas dans le résultat final.

\subsection{Résultats}
\subsubsection{Pré-traitement des données}

La détection des clignements des yeux dans les systèmes de suivi du regard est un problème difficile a gérer entraînant généralement une fausse détection d'un mouvement du regard vers le bas suivi d'un retour à la position initiale. Il en résulte souvent une fausse activation, en particulier pour les éléments en bas du menu. Nos oculomètre Tobii n'ont pas été capables de filtrer ces clignements, ce qui a entraîné de nombreuses fausses activations et un taux d'erreur élevé (environ 10 \% des activations) nous privant de la possibilité de faire une analyse du taux d'erreur. Pour résoudre ce problème, nous avons décidé de supprimer toutes les fausses activations et de nous concentrer uniquement sur les activations des bonnes cibles. D'un point de vue écologique, nous comptons sur une amélioration des oculomètres et, plus particulièrement sur une amélioration du filtrage de ces clignements.

\subsection{Résultats de l'expérience {\em Search and Activate}} 
\subsubsection{Analyse Générale}

\begin{figure}[h!]
\centering
\begin{subfigure}{.2\textwidth}
  \centering
  \includegraphics[width=3.5cm]{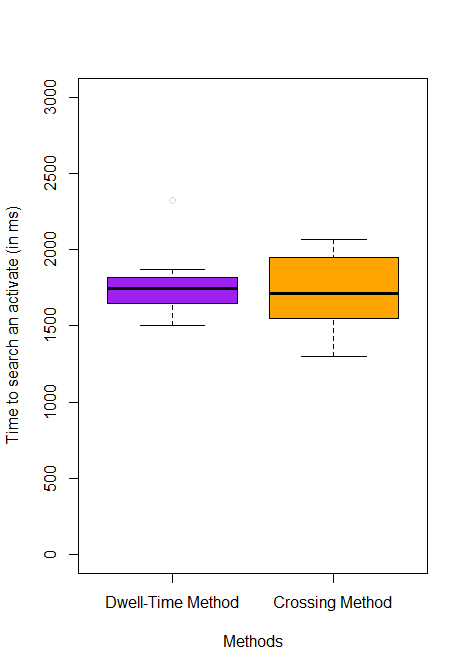}
  \caption{Temps pour sélectionner un élément dont l'utilisateur a pris connaissance.}
\end{subfigure}%
\hspace{0.05\textwidth}\begin{subfigure}{.2\textwidth}
  \centering
  \includegraphics[width=3.5cm]{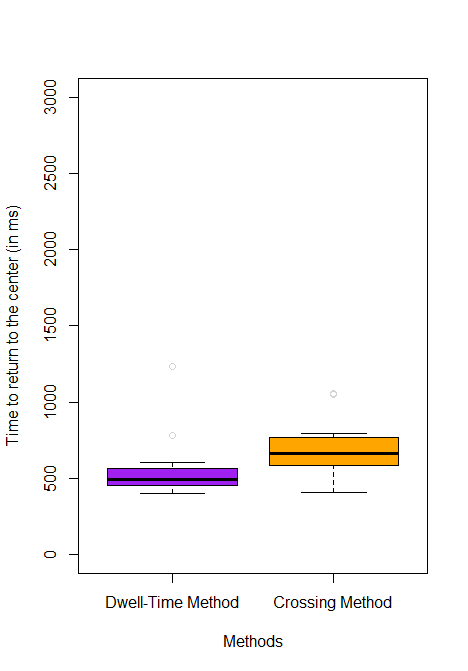}
  \caption{Temps pour retourner au centre du menu.}
\end{subfigure}
\caption{Temps moyens les deux méthodes d'interaction}
\label{mean}
\end{figure}

La Figure \ref{mean} montre la différence de temps entre les deux méthodes d'interaction. Nous pouvons clairement voir que le \verb+return_time+ de la méthode par franchissement est plus long que celui par temps de fixation. Cela peut s'expliquer par le fait que le menu STATU QUO est plus petit que le menu CROSSING. Chaque élément du STATU QUO est alors plus proche du centre du menu. La direction du mouvement des yeux pourrait également avoir un impact (nous savons qu'un mouvement de haut en bas est plus rapide qu'un mouvement de bas en haut par exemple).

Pour l'\verb+activation_time+, on peut voir que le résultat médian des deux méthodes est vraiment similaire. En raison de la grosse variabilité des résultats, il est difficile de dire si l'une ou l'autre méthode est vraiment plus rapide. Le \verb+return_time+ est en moyenne 66,37\% plus rapide que \verb+action_time+ . Ce résultat s'explique par le fait que le temps de recherche d'un élément semble assez long et prend donc plus de temps que l'activation. La longueur du chemin étant en effet la même lors de l'activation d'un élément, et lors du retour au centre à partir de cet élément, il apparaît que la recherche occupe la majeure partie du temps \verb+activation_time+. 
\newline

\begin{figure}[h!]
\centering
\begin{subfigure}{.5\textwidth}
  \centering
  \includegraphics[width=8cm]{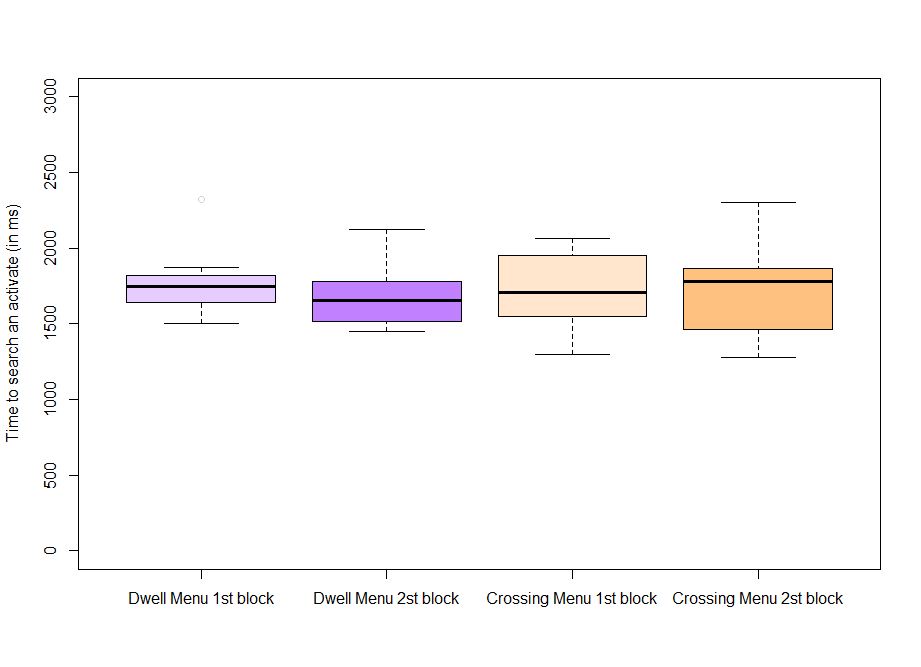}
  \caption{Temps pour sélectionner un élément dont l'utilisateur a pris connaissance.}
\end{subfigure}
\begin{subfigure}{.5\textwidth}
  \centering
  \includegraphics[width=8cm]{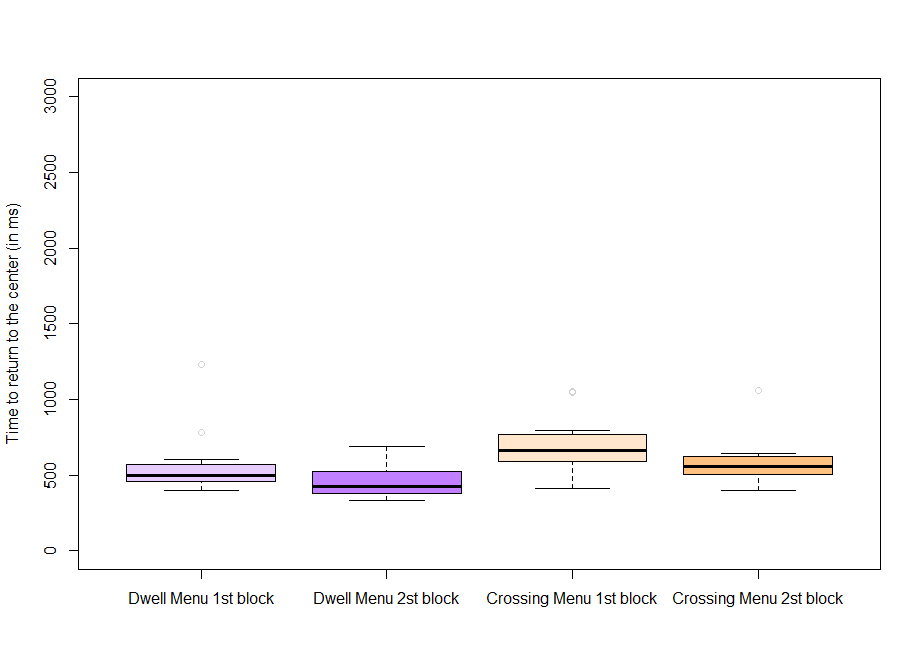}
  \caption{Temps pour retourner au centre du menu.}
\end{subfigure}
\caption{Temps moyen pour chaque bloc en fonction de la méthode de sélection}
\label{mean4}
\end{figure}

La Figure \ref{mean4} représente les temps moyen entre les participants pour les deux blocs de test. Nous voyons clairement que même si le \verb+return_time+ est toujours plus long sur le menu CROSSING comme le montrait la Figure \ref{mean}, il est plus court sur le deuxième bloc pour les deux menus. Le \verb+return_time+ semble avoir un taux d'apprentissage équivalent pour les deux méthodes (19.40\% pour le dwell-time 17.34\% pour la traversée) donc une expérience sur une période plus longue (peut-être plus de sessions) pourrait montrer si cet apprentissage évolue et si l'une ou l'autre méthode est plus rapide pour un utilisateur devenant expert. Pour l'instant, nous pouvons simplement dire que pour un utilisateur novice, la conception plus large de notre menu STATUS QUO semble avoir un impact sur la performance de l'utilisateur.

Pour l' \verb+activation_time+ des deux blocs, en raison de la grande variabilité pour la méthode par franchissement, il est impossible de dire si une méthode est vraiment plus rapide que l'autre. Cependant, on peut noter le fait, contrairement a celle du menu utilisant le temps de fixation, la valeur médiane de l'\verb+activation_time+ pour le menu par franchissement est plus élevée sur le deuxième bloc que sur le premier. Ces résultats pourraient s'expliquer par le fait que les personnes qui n'avaient pas l'habitude d'utiliser l'oculomètre peuvent se fatiguer avec l'utilisation de ces méthodes, ou peuvent tout simplement ne pas avoir subit un entrainement assez long.

\subsubsection{Analyse détaillée}

\begin{figure}[h!]
\centering
\begin{subfigure}{.5\textwidth}
  \centering
  \includegraphics[width=8cm]{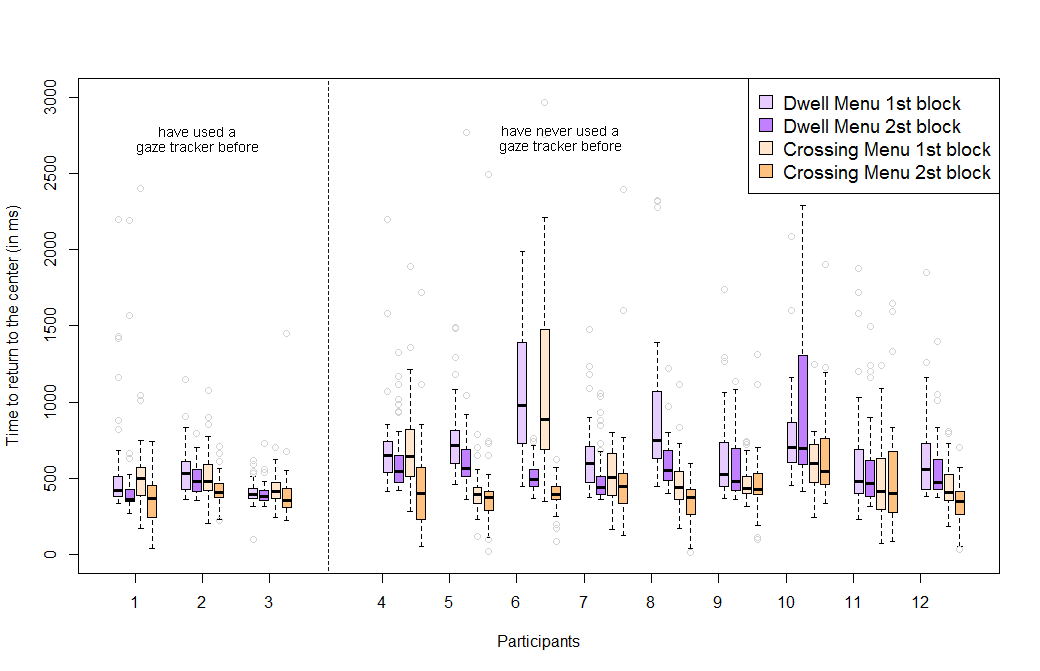}
  \caption{Temps pour sélectionner un élément dont l'utilisateur a pris connaissance.}
\end{subfigure}
\begin{subfigure}{.5\textwidth}
  \centering
  \includegraphics[width=8cm]{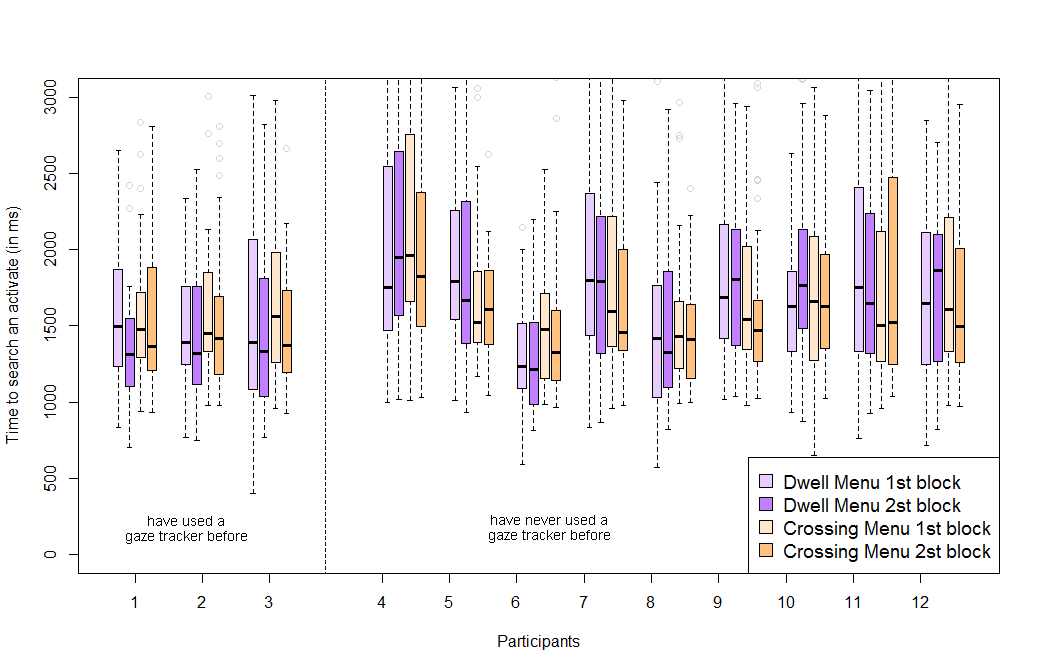}
  \caption{Temps pour retourner au centre du menu.}
\end{subfigure}
\caption{Résultat des temps par utilisateur pour chaque bloc en fonction des méthodes d'interaction}
\label{c}
\end{figure}

En comparant les résultats des blocs 1 et 2 par utilisateur de chaque menu, on peut observer un apprentissage sur le \verb+return_time+, correspondant probablement à un apprentissage de la position du centre du menu. Toutes les valeurs médianes \verb+return_time+ du deuxième bloc de STATUS QUO,a l'exception d'un participant, sont inférieures à celles du premier, pour ce qui est du CROSSING 8 sparticipants ur 12  ont un \verb+return_time+ median plus rapide et plus régulier.

Il en est de meme pour l'\verb+activation_time+ du STATUS QUO, ou 8 participants sur 12 semblent devenir plus régulier (bien que la variabilité soit toujours trop importante pour voir s'il y a une réelle amélioration du temps entre les blocs). 

Le taux d'apprentissage sur le \verb+activation_time+ du CROSSING, lui, semble être très difficile à déterminer, la variabilité des résultats étant très importante. Même si 8 personnes sur 12 ont un temps médian plus rapide sur le deuxième bloc et environ la moitié des 12 participants sont plus réguliers, cet écart trop important ne nous permet pas d'être sûrs de l'amélioration. 
Cela peut signifier que pour devenir plus rapide et plus régulier avec ce type d'interaction, une période d'entraînement plus longue pourrait être nécessaire.

Pour chacun des menus, on distingue neanmois que les utilisateurs ayant déjà utilisé un oculomètre auparavant semblent être en moyenne, plus rapide que les autres.
Pour le STATUS QUO ils sont en moyenne 5,62\% plus rapides le premier bloc et 12,08\% plus rapides sur le second, soit 8,81\% plus rapides en moyenne.
Pour le CROSSING ils sont en moyenne 10.05\% plus rapides qle premier bloc et 20.29\% plus rapides sur le second, soit 15.22\% plus rapides en moyenne.

On voit que le temps médian du \verb+return_time+ sur le menu par franchissement est plus long que celui du menu par temps de fixation dans 10/12 cas, mais là encore, à cause de la grande variabilité, on ne peut déterminer si telle ou telle méthode est vraiment rapide.

Pour l'\verb+activation_time+, la médiane pour l'interaction par franchissement est plus rapide dans la moitié des cas mais ici encore, les variations sont trop extrêmes pour dire si l'une ou l'autre méthode peut être considérée comme plus rapide. 

Ce que nous pouvons aussi remarquer, c'est que les personnes qui ont commencé avec le temps de fixation (1,2,3,4,6 et 10) ont tendance à avoir un meilleur résultat médian par franchissement, tandis que les personnes qui ont commencé avec le franchissement (5,7,8,9,11 et 12) ont tendance à etre plus rapide avec le temps de fixation. Il semble qu'il y ait un transfert d'apprentissage entre les conditions dans les deux sens : STATUT QUO à CROSSING et vice versa.

La différence relative pour le CROSSING entre les personnes qui ont commencé par CROSSING et celles qui ont commencé par STATUS QUO est égale à 11,3\%, contre 0.9\% pour la différence relative pour le STATUS QUO.
Il semble donc que le transfert ne soit pas symétrique car le CROSSING bénéficie davantage d'apprentissage par le STATUS QUO que l'inverse.

Cette différence d'apprentissage pourrait s'expliquer par le fait que la saccade, qui est la principale interaction pour le CROSSING, est entraînée par le STATUS QUO en effectuant la recherche des éléments, mais la fixation, qui est la principale interaction du STATUS QUO, n'est pas entraînée du tout dans le menu CROSSING.

Dans chaque graphique, nous pouvons remarquer qu'il semble y avoir des valeurs aberrantes, elles pourraient être dues à un problème d'étalonnage de l'oculometre ou à une certaine fatigue oculaire pendant l'exercice

\subsection{Résultats de l'expérience d'{\em Activation}}

\subsubsection{Comparison between methods}

Dans les conditions réelles d'utilisation, l'utilisateur n'aura pas à regarder au centre du menu avant de sélectionner un élément. Nous pensons que la suppression de cette action inutile conduira à une activation plus rapide des éléments du menu par franchissement, c'est pourquoi nous avons conçu le tests suivant. Pour cette nouvelle série de tests basés uniquement sur l'activation, nous avons calculé le temps moyen d'activation en alternant l'activation de plusieurs couples de lettres.

\begin{figure}[h!]
  \centering
  \includegraphics[width=0.7\columnwidth]{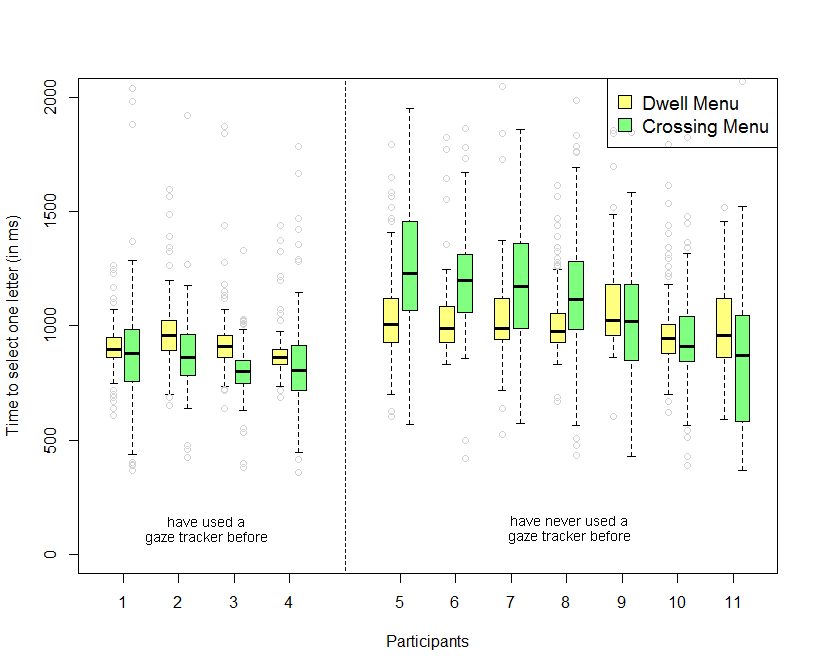}
  \caption{Comparaison du temps par utilisateur pour l'activation d'une lettre avec les deux interactions dans le test de {\em sélection}.}
\label{s}
\end{figure}

Ce que nous pouvons observer avec les résultats de la Figure \ref{s} est le fait que les 4 participants ayant déjà utilisé un oculometre sont plus rapides que la plupart des autres et ont également une variabilité bien plus faible.  Ils sont 15.33\% plus rapides en moyenne avec le temps de fixation et 22.42\% plus rapides avec le franchissement.

En fait, on peut voir que plus un participant est rapide en moyenne, et plus il a tendance à être meilleur avec le menu CROSSING, de l'autre côté, plus le temps médian est lent, plus il semble être plus rapide sur STATUS QUO. Il est important de le noter même si l'écart est trop important pour vérifier que la vitesse moyenne d'un utilisateur influence le fait qu'il aura tendance à être meilleur avec l'une ou l'autre méthode. Il serait intéressant de le tester dans une autre étude. Pour l'instant, comme le montre le ratio entre les temps pour les deux méthodes, il est impossible de une quelconque préférence pour la moyenne des utilisateurs.

En examinant les données en détail, nous avons remarqué que la disposition de quelques lettres peut avoir un impact sur le temps de sélection. Cela explique le grand nombre de valeurs aberrantes que nous pouvons voir dans la Figure \ref{s}. Par exemple, un couple de lettres sur la diagonale du menu par franchissement (A-N ou G-T) sera sélectionné plus rapidement qu'un autre. A l'inverse, pour la sélection par temps de fixation, les couples ayant des lettres proches l'une de l'autre (H-I et J-R), auront pour résultat une sélection plus rapide. Ces informations peuvent être très utiles pour trier les éléments de chacun des menus en vue d'implémentations futures.

\subsection{Retour utilisateur}

Après les phases de tests, les utilisateurs ont pu exprimer librement leurs sentiments sur l'utilisation des deux techniques et à dire pourquoi ils préféraient utiliser l'une ou l'autre.
Parmi les 18 utilisateurs, 8 d'entre eux avaient une préférence pour le menu croisement, 5 autres préféraient le menu par temps de fixation et les 5 derniers n'avaient pas de réelle préférence.
\newline\ newline 
Ceux préférant l'interaction par franchissement ont trouvé le temps de fixation  "frustrant" et "stressant". Pendant la phase de recherche/observation, "l'affichage du retour du temps de fixation était dérangeant", "il attire l'attention" et le temps de fixation est généralement considéré comme "trop court", ce qui conduit à des erreurs de sélection. L'utilisateur doit "garder les yeux en mouvement" ce qui est stressant et fatigant. D'autre part, lors de la phase de sélection, le temps de fixation "trop long", ce qui donne l'impression d'une perte de temps et rend l'utilisateur "incapable de penser à sa prochaine action". Le menu par franchissement crée le sentiment d'une interaction plus rapide, il est moins frustrant car l'utilisateur "n'a pas à se précipiter", il "peut détendre son regard sans activer aucun élément". Le menu par franchissement est plus grand donc il a besoin de plus grandes saccades mais permet de faire moins d'erreurs . Cela peut devenir un problème si l'on veut ajouter plus d'éléments.
\newline \newline
Pour les personnes qui ont préféré l'interaction par fixation, ils ont expliqué que "même elle peut conduire à plus d'erreurs",  l'interaction semble généralement être "plus explicite" que celle par franchissement et peut être comprise plus facilement. De ce fait, l'un des utilisateurs a dit qu'"en s'entraînant davantage avec le franchissement, il croit vraiment qu'il peut devenir plus rapide", l'interaction n'était tout simplement "pas assez explicite" pour lui, cela signifie qu'il lui manquait encore quelques retour visuels, et qu'il a besoin de plus de temps "pour y faire face". Les participants ont également dit que le temps de fixation semble être plus rapide et un peu "plus confortable" car moins de mouvements oculaires sont nécessaires, "surtout lorsque deux lettres sont proches l'une de l'autre". 

\section{Conclusion}


Dans cet article, nous avons proposé un menu circulaire par franchissement utilisable avec le regard. Nous avons montré que certains utilisateurs appréhendaient parfaitement cette interaction qui permet aux utilisateurs de reposer le regard dans endroits sans zone de franchissement et élimine le besoin de faire une pause dans l'interaction.
Nous avons suggéré que l'ajout d'un algorithme estompant les effets négatifs entraînés par les clignements des yeux devrait réduire le taux d'erreur dans notre implémentation de l'interaction par franchissements.  À l'heure actuelle, l'absence de mise en œuvre de cet algorithme peut perturber l'utilisateur et est certainement à l'origine de l'interaction plus lente pour les utilisateurs novices, c'est-à-dire ceux qui utilisent pour la première fois des traceurs oculaires. 

Cette étude peut être considérée comme complémentaire au projet {\em Gazeplay}\footnote{\url{http://www.gazeplay.net/}}\footnote{\url{https://github.com/schwabdidier/GazePlay}} \cite{schwab}, créé en 2016 par Didier Schwab. L'idée principale de Gazeplay est de proposer un logiciel et une plateforme de jeu à bas prix pour les enfants en situation de polyhandicap, utilisables avec un oculomètre. Le coût des technologies concurrente étant élevé, ce projet est compatible avec des appareils moins chers comme le {\em Tobii 4C} et offre une plateforme logicielle libre et gratuite non seulement pour les familles mais aussi pour les centres médicaux. En plus d'être utile pour rendre les enfants polyhandicapés capables de jouer "librement", la création de cette plateforme vise à enseigner et à améliorer les capacités d'interaction visuelle et de cognition de ces enfants. Nous prévoyons ainsi d'intégrer l'interaction par franchissement étudiée dans cet article dans certains jeux de Gazeplay afin de la tester avec les enfants polyhandicapés qui sont, avec leurs aidants, parmi les personnes susceptibles d'être les plus intéressées par cette nouvelle technique d'interaction.

\bibliographystyle{ACM-Reference-Format}
\bibliography{references}



\end{document}